\documentclass[letterpaper,floatfix,preprint,superscriptaddress,longbibliography,amsmath,amssymb,aps,pra]{revtex4-2}
\usepackage{graphicx,setspace}
\usepackage{braket}
\usepackage{filecontents}

\newcommand{\unit}{\mathop{}\!\mathrm}
\newcommand{\Ham}{\mathcal{H}}

\begin{document}
	
\title{Quantum-assisted distortion-free audio Signal Sensing}%
	
\author{Chen Zhang}%
\email[]{c.zhang@pi3.uni-stuttgart.de}
\affiliation{3rd Institute of Physics, University of Stuttgart, Pfaffenwaldring 57, Stuttgart 70569, Germany}
	
\author{Durga Dasari}%
\email[]{d.dasari@pi3.uni-stuttgart.de}
\affiliation{3rd Institute of Physics, University of Stuttgart, Pfaffenwaldring 57, Stuttgart 70569, Germany}
	
\author{Matthias Widmann}%
\affiliation{3rd Institute of Physics, University of Stuttgart, Pfaffenwaldring 57, Stuttgart 70569, Germany}
	
\author{Jonas Meinel}%
\affiliation{3rd Institute of Physics, University of Stuttgart, Pfaffenwaldring 57, Stuttgart 70569, Germany}

\author{Vadim Vorobyov}%
\affiliation{3rd Institute of Physics, University of Stuttgart, Pfaffenwaldring 57, Stuttgart 70569, Germany}

\author{Polina Kapitanova}%
\affiliation{Department of Physics and Engineering, ITMO University, Saint Petersburg 197101, Russia}
	
\author{Elizaveta Nenasheva}%
\affiliation{Giricond Research Institute, Ceramics Co. Ltd., Saint Petersburg 194223, Russia}
	
\author{Kazuo Nakamura}%
\affiliation{Hydrogen and Carbon Management Technology Section, Hydrogen and Carbon Management Technology Strategy Department, Tokyo Gas Co., Ltd., Yokohama 230-0045, Japan}
	
\author{Hitoshi Sumiya}%
\affiliation{Advanced Materials Labotratory, Sumitomo Electric Industries, Ltd., Itami 664-0016, Japan}
	
\author{Shinobu Onoda}%
\affiliation{Takasaki Advanced Radiation Research Institute, National Institutes for Quantum Science and Technology, Takasaki 370-1292, Japan}
	
\author{Junichi Isoya}%
\affiliation{Faculty of Pure and Applied Sciences, University of Tsukuba, Tsukuba 305-8573, Japan}
	
\author{Jörg Wrachtrup}%
\affiliation{3rd Institute of Physics, University of Stuttgart, Pfaffenwaldring 57, Stuttgart 70569, Germany}
	
	
\begin{abstract}
	Quantum sensors are keeping the cutting-edge sensitivities in metrology. However, for high-sensitive measurements of arbitrary signals, limitations in linear dynamic range could introduce distortions when sensing the frequency, magnitude and phase of unknown signals. Here, we overcome these limitations with advanced sensing protocol that combines quantum phase-sensitive detection with the heterodyne readout. We present theoretical and experimental investigations using nitrogen-vacancy centers in diamond, showing the ability to sense radio signals with a 98 dB linear dynamic range, a 31 pT/Hz$^{1/2}$ sensitivity, and arbitrary frequency resolution. Further, we perform the quantum-assisted distortion-free audio signal (melody, speech) sensing with high fidelity. The methods developed here could broaden the horizon for quantum sensors towards applications in telecommunication, where high-fidelity and low-distortion at multiple frequency bands within small sensing volumes are required.
\end{abstract}
	
\maketitle
\section{Introduction}
	
Quantum sensors are setting new frontiers of sensing techniques with their extraordinary performances in sensitivity and stability \cite{RN1,RN2,RN3,RN4,RN5}.
These techniques rely on either measuring the line-shift of spin or atomic transition frequencies or reading out the relative populations of the occupied energy levels using interferometric methods \cite{RN6, RN7}.
In most cases, there are trade-off relations between the sensitivity and other features in metrology \cite{RN9}.
For example, a high-sensitive measurement acquired by detecting the transition line shift requires a narrow linewidth, which, on the other hand, will limit the dynamic range.
Interferometric measurements detect a sinusoidal response, and linearity is only achieved when the phase signal is in a small dynamic range.
It sets a massive limitation on the sensitivity when sensing an unknown signal that gets measured beyond this linear regime, for example, when the working point of the sensor is at the maxima or minima of the sinusoidal signal response.
Thus, it becomes a bottleneck for high sensitivity measurements that are required in many cutting-edge applications.
Operating within the linear dynamic range (LDR) can be crucial for reconstructing unknown signals.
One way to directly extract the phase factor, which is linear to the physical quantity to be detected, is to use phase-sensitive detection known as the classical lock-in technique.
In this work, using a modified sensing scheme that introduces an external readout phase modulation, we acquire the target quantum phase signal after demodulation.
Therefore, the LDR is no longer limited to the small-angle approximation.
Hence our method combines large dynamic range with maximum sensitivity.

Nitrogen-vacancy (NV) centers in diamond have been at the forefront in performing high-sensitive measurements of various physical quantities, viz., magnetic and electric field, temperature, and strain distributions internal and external to diamond \cite{RN10,RN11,RN12,RN13, RN41}.
The NV magnetometry has been performed under bias fields ranging from zero-field to a few Tesla, and for sensing signals with frequencies ranging from DC to a few GHz \cite{RN14,RN15,RN16,RN17}.
While dynamical-decoupling techniques are usually employed for high sensitivity \cite{RN10,RN18,RN19}, arbitrary frequency resolution can be achieved with the quantum heterodyne (Q-dyne) detection \cite{RN20,RN21}.
However, both methods suffer from a limited LDR when they are applied to measure arbitrary signals.

For high dynamic range measurements, a closed-loop frequency-locking scheme together with optically detected magnetic resonance (ODMR) can be used to track resonance frequency shifts \cite{RN22}.
However, this scheme cannot be used for ac field measurements in combination with interferometric methods, if the signal frequency is higher than the readout sampling frequency.
Phase-estimation algorithms can effectively improve the LDR in Ramsey measurements by varying the sequence with adaptive feedback schemes \cite{RN8,RN42}.
However, for the case of ac sensing schemes e.g. Hahn-echo, as varying the sequence itself will change the sensor response to the ac signals, such methods become less applicable. 
Therefore, a technique is still missing, that addresses the LDR while maintaining high sensitivity and frequency resolution, for example, in sensing arbitrary radio-frequency fields within a broad bandwidth.

Sensing radio-frequency signals by electric-field sensors, either conventional electronic receivers or the Rydberg atom sensors, need antennas to collect and guide the electric signals towards the sensors \cite{RN23,RN24,RN25,RN26}.
Although the receivers can be highly integrated, the dimension of antennas can scale to meters due to the signal wavelength.
The size becomes critical when there is limited space for the sensor, for example, in a satellite. In this regard, quantum magnetometers can be very attractive due to their small sensing volume and high sensitivity \cite{RN27}.
A flux concentrator can be used as a substitute to conventional antennas for obtaining high signal gain. Independent of the signal wavelength, the dimensions of such flux concentrators can be as small as a few centimeters \cite{RN28,RN29}.

In this paper, we demonstrate the quantum-assisted distortion-free audio signal sensing with NV center ensembles in diamond using a quantum-phase-sensitive detection (QPSD) technique combined with heterodyne readout.
Firstly, we introduce the QPSD technique, which can provide an extended LDR in interferometry measurements by using two synchronized driving fields.
Then, we present the heterodyne readout, which can interpret e.g. radio signals to get frequency information.
Taking advantage of the bandwidth of the Hahn-echo sequence and the frequency comb induced by the continuous sampling, we demonstrate measurements of audio signals around 10 kHz, beyond the coherence limit without losing sensitivity.
Finally, we present arbitrary radio signal measurements with a LDR of 98 dB at a sensitivity of 31 pT/Hz1/2.
A piece of melody and a speech are encoded as magnetic field signals and measured by the NV magnetometer.
By using the sensor as a quantum radio receiver, we demonstrate the application potentials for areas such as quantum-assisted telecommunication and unknown signal exploration. 

\section{Results}
\subsection{Quantum Phase Sensitive Detection}
\begin{figure}[t]
	\includegraphics[scale=1]{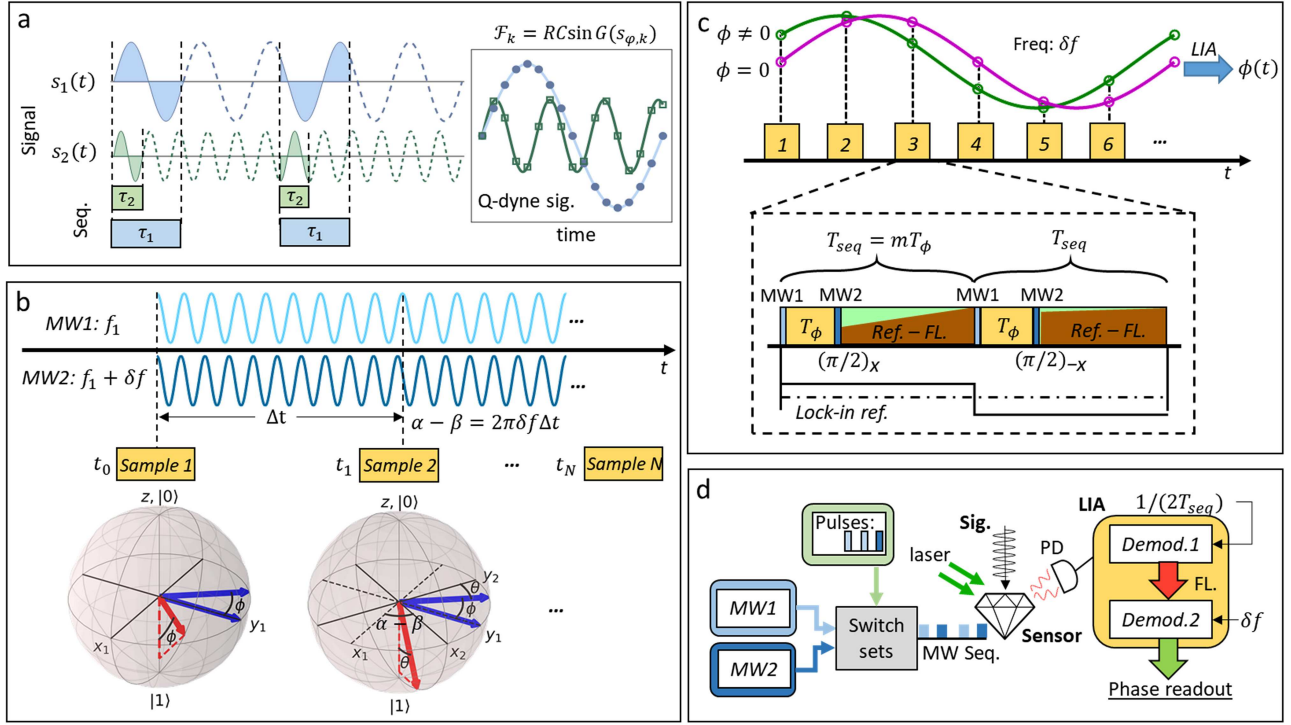}
	\caption{\label{fig1} Phase-sensitive NV magnetometry 
		(A) Continuous sampling induced phase reviving signals, known as the quantum heterodyne (Q-dyne) detection.
		The phase reviving frequency changes with the external field and cannot be used for phase-sensitive detection.
		The signal responses $G(s_\phi, k)$ should be small to ensure measurement linearity.
		$R$ is the detected photon rate, and $C$ is the detected signal contrast.
		(B) Unlike the Q-dyne detection, the quantum phase-sensitive detection (QPSD) is based on the rotating frame modulation induced by the evolving phase difference of the two driving MW fields.
		Two frequency-offset MWs acquire a phase difference of $\alpha-\beta=2\pi\delta f\Delta t$ after the sampling time interval $\Delta t$.
		In the Bloch sphere picture, it can be understand as the MW2 defined rotating frame $x_2y_2z$ rotates with rate of $\delta f$ referring to the MW1 defined rotating frame $x_1y_1z$.
		The acquired quantum phase is $\theta=2\pi\delta f \Delta t-\phi$ at sample 2.
		Through the quantum phase modulation, the acquired readout representing the Bloch vector projections is as shown in
		(C), where we present the measurements of the quantum phase $\phi=0$ and $\phi\neq0$.
		By demodulating the acquired signal with a lock-in amplifier, we can get the phase values.
		The dashed box shows the measurement sequences we applied in experiments.
		Except for the last $\pi/2$ pulse applied with MW2, all the other driving pulses are generated by MW1.
		The fluorescence signal is demodulated at the frequency of $1/(2T_{seq})$ to get a fluorescence intensity readout for a sample.
		The QPSD readout is acquired with the demodulation at $\delta f$.
		(D) Schematic of the experiment. NV centers ensemble in diamond is used to perform the QPSD readout.
	}
\end{figure}
In interferometric measurements, the quantum phase to be detected is usually converted to a quantum state population difference, resulting in a sinusoid readout \cite{RN10}.
A way to extract the phase factor from the sinusoidal readout is to modulate the phase with a specific frequency and perform phase-sensitive modulation.
Such a quantum phase modulation can be introduced by using the difference between the quantum phase of the sensor to an external oscillator.
The Q-dyne method uses such a strategy for resolving frequency of signals better than the relaxation time of the sensor, as shown in Fig. \ref{fig1}a \cite{RN20,RN21}.
However, it cannot be used for phase-sensitive detection because the Q-dyne frequency is also what to be resolved and an extra modulation is still needed \cite{RN21}.
Another way to introduce such a phase modulation is to use the frequency offset between two different driving fields of the sensor \cite{RN30,RN31}.
These driving fields define two rotating frames, and the evolution of the spin as seen from one rotating frame will lead to a quantum phase modulation due to the relative rotation of the two frames, i.e., rotating frame modulation.
The modulation frequency only depends on the frequency difference of the two driving fields, as shown in Fig. \ref{fig1}b and c.
By performing multiple measurements within a modulation cycle and upon using lock-in detection, we will achieve phase-sensitive detection.
Below we mathematically describe this relative evolution of the sensor under such interferometric measurement with two-frequency driving fields.

Aligning an external field $B_0$ along the NV axis, we use the two-level subspace of the NV ground triplet in the derivation.
Thus, the Hamiltonian of the system can be written as:
\begin{equation}
	\Ham = \omega_0 S_z + \gamma_e B_1 \cos\left(2\pi f_1 t+\alpha\right)S_x,
\end{equation}
where $\omega_0$ is the transition frequency of the two-level subspace, $B_1$ is the oscillating magnetic field perpendicular to the NV axis, $f_1$ and $\alpha$ are the frequency and phase of the driving field, and $\gamma_e$ is the electron gyromagnetic ratio.
In the rotating frame defined by the resonance frequency, the time-dependent Hamiltonian is
\begin{equation}
	\Ham_1' = \Omega_1 \cos\left(\delta\omega_1+\alpha\right)S_x + \Omega_1 \sin\left(\delta\omega_1 t+\alpha\right)S_y,
\end{equation}
where $\delta\omega_1=2\pi(f_0-f_1)$, and $\Omega_1=\gamma_e B_1/2$ is the Rabi frequency introduced by MW1.
In interferometry measurements, a $\pi/2$  pulse prepares the spin state from the polarized state to an equalized population, and another $\pi/2$ pulse projects the quantum phase as a population difference after the sensing procedure.
We use the second driving field, MW2, to offset the frequency of the second $\pi/2$ pulse.
$\delta \omega_2$,$\beta$ and $\Omega_2$ are used to denote the frequency offset, Rabi frequency, and phase of MW2.
After this, the measured spin-expectation value is
\begin{equation}
	\braket{S_z} = \sin\left[\phi+\frac{\pi}{2}\left(\frac{\delta\omega_1}{\Omega_1}-\frac{\delta\omega_2}{\Omega_2}\right)+\alpha-\beta\right],
\end{equation}
where $\phi$ is the acquired quantum phase which contains the information we want to measure, both of the MWs are near-resonant with $\delta\omega_1\ll\Omega_1$, $\delta\omega_2 \ll \Omega_2$.
Therefore, the off-resonant term can be neglected, and the phase difference term $\alpha-\beta$ will evolve with time so that there is
\begin{equation}
	\braket{S_z} \approx \sin \left(\phi+2\pi\delta f\cdot t\right),
\end{equation}
where $\delta f$ is the frequency difference of the two MWs.

The above result can be seen as a modulation of the rotating frame itself.
As schematically shown in Fig. \ref{fig1}b (left Bloch sphere), we can assume that the two driving fields have the same phase at the duration of the second $\pi/2$ pulse, and this defines an instantaneous rotating frame with coordinates $x_1 y_1 z$.
Thus, the readout is similar to that of the regular Ramsey interferometry using a single driving field.
After an interval of $\Delta t$, MW2 develops a phase difference of $2\pi \delta f \Delta t$. Since the quantum phase is finally measured by MW2, the Bloch vector rotates in the new instantaneous rotating frame with coordinates $x_2 y_2 z$, as shown in Fig. \ref{fig1}b (the right Bloch sphere).
The rotating frame defined by MW2 rotates continuously around the $z$-axis with the frequency of $\delta f$.
Due to this, the fluorescence readout also modulates in a sinusoidal fashion, as shown in Fig. \ref{fig1}c.
While the readout signal frequency depends on $\delta f$ and its amplitude depends on the signal contrast, the initial phase, $\varphi$, is linear to the field to be measured.
Through the external modulation induced by the MWs, the working point of the sensor evolves in the entire phase range, which ensures the LDR of the initial phase measurement covering $[-\pi, \pi]$.
By fitting or demodulating the fluorescence signal, we can resolve the changing of the phase factor $\varphi$ between each modulation cycle and find measurement linearity for the external field.

The measurement sequence we applied in the experiment is depicted in Fig. \ref{fig1}c, in which $T_\phi$ is the field sensing time, $T_{seq}$ is the sequence length of one measurement, and we use a second measurement with the final pulse changed to $(\pi/2)_{-x}$.
As a result, the fluorescence signal is modulated with a frequency of $f_s=1/(2T_{seq})$, which is also the sampling frequency of the fluorescence readout.
The demodulation of the fluorescence signal, denoted by \textit{Demod. 1} in Fig. \ref{fig1}d, has a readout bandwidth $f_s/2$ set by the Shannon sampling theorem.
The readout is further demodulated by another demodulator of the lock-in amplifier (LIA), denoted as \textit{Demod. 2}.
Upon measuring $N$ samples of the fluorescence readout, the bandwidth of the phase readout is narrowed down to $f_s/(2N)$.
These measurements are schematically shown in Fig. \ref{fig1}d.

The sensitivity of such measurements can be derived based on the fitting of the $N$ samples in the measurement time of $N\cdot2T_{seq}$.
The minimum detectable phase is derived as
\begin{equation}
	\delta\phi=\frac{2}{\sqrt{N}}\frac{1}{C\sqrt{\mathcal{N}}},
\end{equation}
where $C$ is the fluorescence signal contrast, $\mathcal{N}$ is the detected photon counts in each measurement. The sensitivity to external magnetic field, however, is still subject to the applied MW sequence, can be derived as
\begin{equation}
	\eta=\frac{2}{\gamma_e|G(\omega)|C}\sqrt{\frac{2T_{seq}}{{\mathcal{N}}}},
\end{equation}
where $|G(\omega)|$ is the MW filter function which is usually used to describe the transfer function of such a sensor from magnetic field to quantum phase.
In comparison to the conventional fluorescence readout, the sensitivity of QPSD readout deteriorates by a factor of $\sqrt{2}$.
Details about the sensitivity derivation can be seen in the Supplementary Materials.

\begin{figure}[t]
	\includegraphics[scale=1]{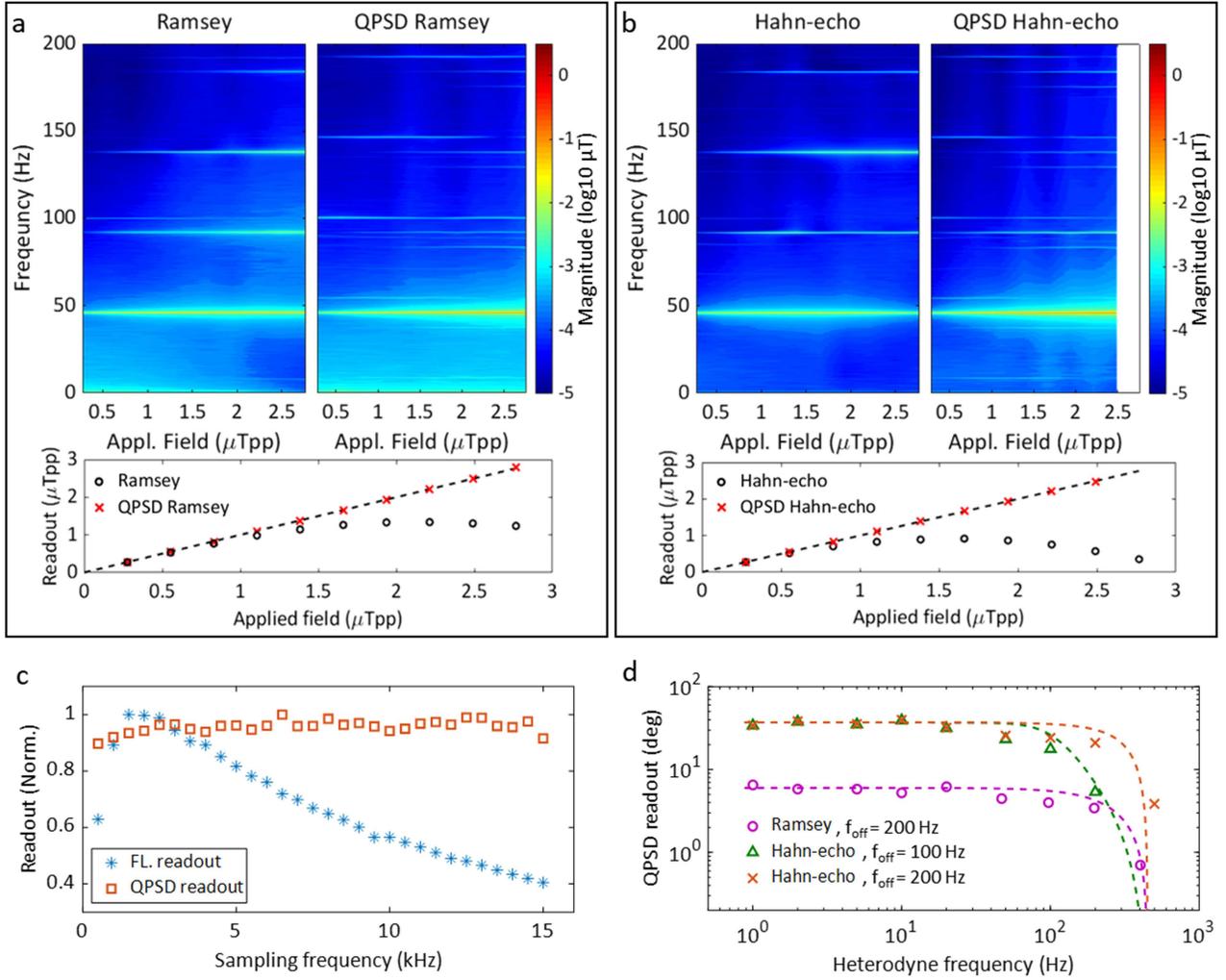}
	\caption{\label{fig2} Sensing performance of the QPSD.
		(A) Spectra and linearity comparison of the normal Ramsey readout and the QPSD readout.
		We apply $T_\phi=6.25\mu s$ in both measurements.
		The applied peak-to-peak field and the readout are plotted showing the linearity of the measurements.
		(B) Spectra and linearity comparison of normal Hahn-echo readout and the QPSD readout.
		The applied phase accumulation time $T_\phi=12.5\mu s$.
		Thus, the Hahn-echo measurements performs a higher sensitivity but smaller dynamic range than the Ramsey measurements.
		(C) Signal response to different sampling frequencies.
		The measurements use the same calibration field, and the readouts are normalized to be plotted in the same vertical axis. 
		(D) Measurement bandwidth.
		Ramsey and Hahn-echo sequences are applied to measure test fields at different frequencies with the same magnitude.
		The heterodyne frequency responses are limited in bandwidth by the cut-off frequency of the LIA.
	}
\end{figure}

In Fig. \ref{fig2}a and b, we compare the regular interferometry (single driving field) and with the measurements obtained from the QPSD readout described above.
The strength of the applied external ac fields ranges from 0 to 3 $\unit{\mu T}$.
For Ramsey measurements, the applied fields are at a frequency of 46 Hz, and we use a field sensing time $T_{\phi,Ramsey}=6.25 \unit{\mu s}$.
For Hahn-echo measurements, we use external fields at 80 kHz+46 Hz and the field sensing time $T_{\phi,Hahn}=2T_{\phi,Ramsey}$.
The test fields are sent to the diamond by a calibrated loop antenna.
The signal readout of the regular interferometry measurements is proportional to $\sin(\phi)$, where $\phi \propto \gamma_e B$ is the accumulated quantum phase.
The response is linear only when $\phi$ is small, limiting the dynamic range.
Thus, the regular Ramsey and Hahn-echo readout quickly saturate due to this limited LDR.
We plot the Fourier transform of the readout signals also in the figures.
The harmonics of the 46 Hz signal rise significantly in the fluorescence readout spectral due to the saturation induced by the limited LDR, compared to the QPSD readout which shows the linearity over the entire field range.
The high-order harmonics of the signal detected by the QPSD readout are small and mainly arise from the function generator.
In the measurements, one could see the linewidth broadening induced by the increasing signal power.
The peak at 100 Hz, which is consistently seen in both the Ramsey and Hahn-echo measurements, comes from the electronics instrumentation.
Other side peaks seen near the original signal frequency in the QPSD readout spectra are due to the mixing of the 100 Hz power line harmonics and the 92 Hz signal harmonics in the LIA.

Besides the LDR, the method also demonstrates measurement robustness to changing of $T_{seq}$. The motivation of using different $T_{seq}$ is to get different sampling frequencies as well as measurement bandwidth.
Signal responses to different sampling frequencies, i.e., different $1/2T_{seq}$, are plotted in Fig. \ref{fig2}c.
Characterized by the same test field, fluorescence readout shows varying signal responses over the sampling frequency range, while the QPSD readout almost stays at the same level because the measured phase factor only changes with the external field and the sensing time $T_\phi$.
It also indicates that the QPSD readout does not change with varying of fluorescence signal contrast, which is affected by the low spin polarization rate when $T_{seq}$ is short in the regime of low excitation laser power.

The measurement bandwidth of the QPSD readout is shown in Fig. \ref{fig2}d, where the signal responses to different test field frequencies are plotted.
The plotted values are the magnitudes at the corresponding frequencies in the Fourier transform of the QPSD readout.
For the measurements based on the Hahn-echo sequence, we detected the heterodyne signal for the ac fields.
The applied sequence length, $T_{seq}=100 \unit{\mu s}$, gives the referencing frequency $f_s=$5 kHz for \textit{Demod. 1}.
We apply the second driving field with the frequency offset at $\delta f = 500$ Hz to have $N=10$ samples in a modulation cycle.
Due to this, there will be flexibility in deciding the single measurement bandwidth by setting the time constant of \textit{Demod. 2}.
We choose different settings corresponding to the cut-off frequency of the filter at 100 Hz and 200 Hz.
Finally, one can conclude that the rotating frame modulation provides QPSD readout magnetometry that has enhanced LDR and robustness in a flexible bandwidth.
As we show below, this can be used for measurement of arbitrary fields with low distortion.
	
\subsection{Frequency Offset Heterodyne Readout}
Heterodyne readout has been used to improve the frequency resolution remarkably in nuclear magnetic resonance spectroscopy.
It is also a way to achieve high precision microwave sensing \cite{RN32,RN33,RN34}.
High-order dynamical decoupling sequences are used to narrow the spectral linewidth by decoupling the sensor response from unwanted signal frequencies \cite{RN20,RN21}.
Here arises a trade-off between the measurable signal bandwidth and fidelity.
High-order dynamical decoupling can ensure a high sensitivity but only allows to measure signals within the narrow bandwidth defined by the sequence.
On the other hand, the lower limit on the detectable signal frequency is set by the decoherence time of the sensor.
Here, we will use the Hahn-echo sequence in combination with the QPSD readout to measure signals at frequencies that go beyond the coherence time of the sensor.

\begin{figure}[t]
	\includegraphics[scale=0.95]{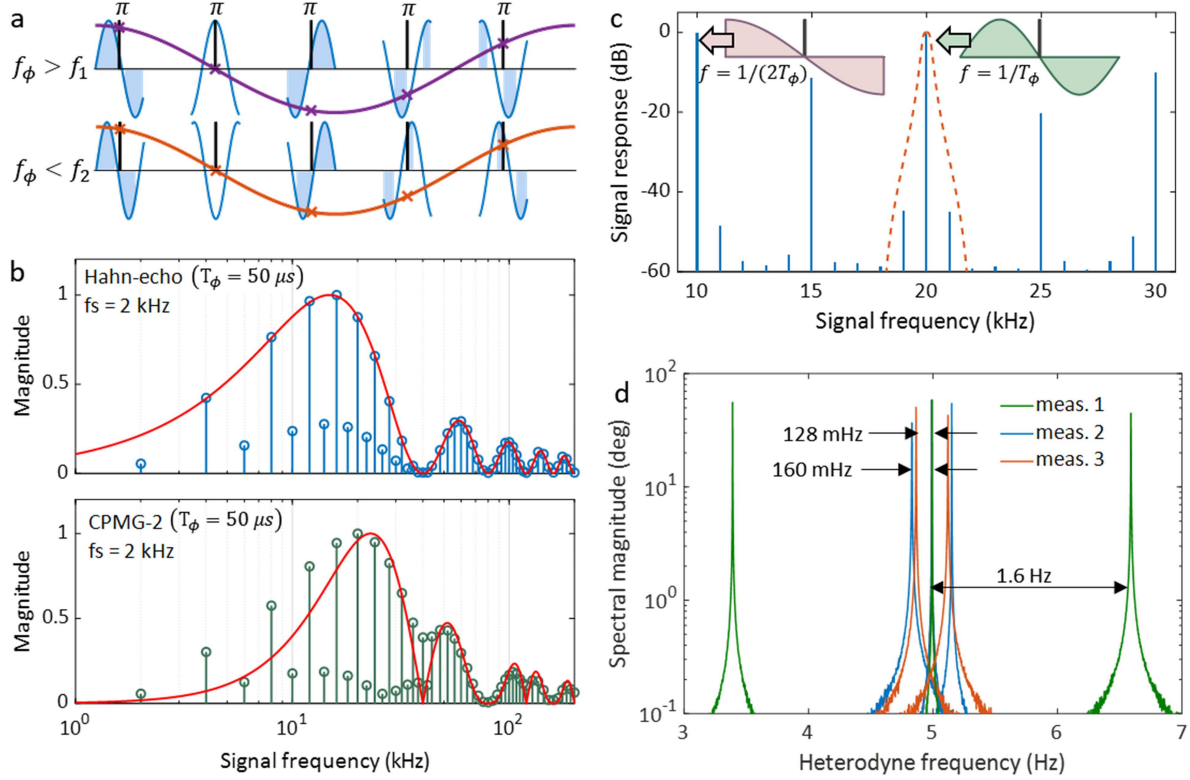}
	\caption{\label{fig3} Frequency Offset Heterodyne readout.
	(a) Hahn-echo sequence is used for this demonstration.
	The detected phase of the ac signal is locked by the sequence when the frequency is $f_{ac}=f_\phi$. Otherwise, a heterodyne signal of $\left|f_{ac}-f_\phi\right|$ can be measured.
	The colored regions mark where the quantum phase is accumulated, while phase accumulations at the other areas are canceled in the spin evolution.
	The figure shows identical heterodyne signals due to $f_\phi-f_1=f_2-f_\phi$.
	(b) We apply ac fields at different frequencies with an offset of 5 Hz to the sensor so that 5 Hz peaks can be detected as the signal response.
	The signal frequency response of the Hahn-echo sequence and CPMG-2 sequence are plotted after normalization, respectively.
	In both measurements $T_\phi=50 \unit{\mu s}$, and the sampling frequency is 2 kHz according to the applied sequence length. The red lines are the filter functions in theory.
	(c) Signal frequency response of Hahn-echo measurements with $1/T_{seq}=10$ kHz.
	The dash line indicates the filter introduced by the lock-in amplifier.
	(d) Sequence dependency of the frequency resolution.
	In meas. 1, a 20.005 kHz filed is applied and measured by sequences with $T_\phi=50 \unit{\mu s} \pm 4 \unit{ns}$, and $T_{seq}=20T_\phi$.
	In meas. 2, we keep $T_\phi$ unchanged, and offset $T_{seq}$ with $\pm$4 ns.
	In meas. 3, the frequency of the applied field is changed to 16.005 kHz while the other parameters are the same as meas. 2.
	}
\end{figure}

In Qdyne, the sampling time usually satisfies $T_{seq}\neq mT_\phi$ so as to get the heterodyne signal \cite{RN21}.
The frequency of this heterodyne signal depends on the timing offset.
Here, we choose the measurement sampling time $T_{seq}=mT_\phi$ to obtain the heterodyne readout depending on the signal frequency offset from $1/T_\phi$.
As a result, the detected phase of signals at frequencies of $n/T_\phi$ is locked by the sequence, where $n$ can be a random integer.
On the other hand, the frequency offset of signals can also introduce phase revivals, i.e. frequency offset heterodyne signal, as shown in Fig. \ref{fig3}a.
The detected heterodyne frequency would be the exact offset of the signal frequency to $1/T_\phi$.

The frequency offset heterodyne readout is modeled based on the MW sequence filter \cite{RN35,RN36}. 
Sampling happens in each time interval of $\left[NmT_\phi,(Nm+1)T_\phi\right]$, where $N\in\mathbb{Z}$.
For a random ac signal component $B_{ac}(t)=B(\omega)e^{-i[\omega t+\varphi(\omega)]}$ and a measurement with the MW $\pi$-pulse number of $n$, we can get the accumulated quantum phase as (derivation see in the Supplementary Materials)
\begin{equation}
	\phi_r(N)=\left|G_n(\omega)\right|e^{i\left(-\frac{\omega T_\phi}{2}-\frac{P}{2}\pi\right)}\gamma_e B(\omega) e^{-i\varphi(\omega)}e^{-i \omega NmT_\phi}
	\label{eq_arb_sig},
\end{equation}
where $N$ denotes the sampling timestamp, $G_n(\omega)=\left|G_n(\omega)\right|e^{i\left(-\frac{\omega T_\phi}{2}-\frac{P}{2}\pi\right)}$ is the MW filter function, $P=1$ when the $\pi$-pulse number $n$ is odd and $P=2$ when $n$ is even.
Particularly, when $n=1$ i.e. Hahn-echo sequence is applied, the filter function satisfies $|G_1(2\pi/T_\phi)|=|G_1(\pi/T_\phi)|$.
In principle, measurements of signals at a wide frequency range is feasible by choosing the appropriate $T_\phi$ in Hahn-echo measurements.
For example, by using $T_\phi<1 \unit{\mu s}$, one can achieve detection of signals at frequencies higher than 1 MHz.
It is more challenging to measure a signal at a lower frequency, such as a signal at 10 kHz, for the reason that a longer $T_2$ is required.
With the property described above, it is feasible to use $T_\phi=50 \unit{\mu s}$ rather than $T_\phi=100 \unit{\mu s}$ to achieve the measurement with a better sensitivity due to the higher signal contrast when $T_\phi$ is smaller.
For diamonds which have NV center ensembles with $T_2<100 \unit{\mu s}$, the property makes it feasible to measure signals at the frequencies lower than $1/T_2$ beyond the coherence limit.

Given a reference frequency $\omega_{ref} = k\omega_s, k\in\mathbb{N}$, where $\omega_s=2\pi/(mT_\phi)$ and $\omega \in \left(\omega_{ref}-\omega_s/2,\omega_{ref}+\omega_s/2\right)$, the evolving phase factor can be rewritten as $e^{-i\omega NmT_\phi}=e^{-i\omega_H t}\delta(t-NT_s)$, where $\omega_H=\omega-\omega_{ref}$ is the heterodyne frequency, $\delta(t)$ is the Dirac function, and $T_s=mT_\phi$ is the sampling period.
Thus, the readout signal turns to be
\begin{equation}
	\phi_r(t)=G(\omega)\sum_{N=-\infty}^{\infty}\gamma_e B_H(t) \delta(t-NT_s),
	\label{eq_phi_r_t}
\end{equation}
where $B_H(t)=B(\omega)e^{-i(\omega_Ht+\varphi)}$ contains all the information from the origin signal to be measured.
As discussed in previous section, the quantum phase readout bandwidth is limited by the cut-off frequency $f_c$ of the filter of LIA. For any signal with a frequency range of $[(k-1)f_s+f_c, (k+1)f_s-f_c]$, aliasing can be filtered.
Although a smaller $f_c$ makes the measurement bandwidth narrower, it ensures signals that in a larger frequency range can be detected without aliasing.
By changing $T_\phi$ together with $T_{seq}$, we can resolve a spectrum in multiple frequency bands with a series of sequences.

We present two specific examples of the measured frequency responses by using the Hahn-echo and CPMG-2 sequence.
We plot both the theoretical MW filter function and the experimentally measured signal responses together in Fig. \ref{fig3}b.
The field sensing time for both experiment and theory calculations is set to be $T_\phi=50 \unit{\mu s}$.
In the experiments, we measured the amplitudes of the frequency offset heterodyne signals with $T_{seq}=250 \unit{\mu s}$, i.e., the magnetic field sampling rate is 4 kHz.
Due to this reason, the measured MW filters are combed with a frequency distance of 4 kHz. Aliasing signals exist between the main lobes at a distance of 2 kHz, because the readout sampling frequency is $f_s=2$ kHz.

In order to measure signals that distribute in larger bandwidth, we can increase the sampling frequency, for example, to $f_s=5$ kHz.
The spectrum is plotted in Fig. \ref{fig3}c in decibel, from which one can see that magnitudes are the same at 10 kHz and 20 kHz, i.e., $1/(2T_\phi)$ and $1/T_\phi$ as discussed in the derivation.
The insets of Fig. \ref{fig3}c depict the signals that the quantum sensor detects during $T_\phi$ at the two frequencies.
In this measurement, the bandwidth limited by the filter of the LIA is at 200 Hz, i.e., the single measurement bandwidth is 400 Hz, and the detectable signal frequency range is 9600 Hz.

We notice that a single measurement cannot tell if the ac field frequency offset is positive or negative from the heterodyne readout.
Additional measurement is needed to distinguish the direction of the frequency offset.
By adding a difference to the phase accumulation time $T_\phi$ as well as the sequence time, we can change the reference frequency $\omega_{ref}$ to get a different heterodyne frequency.
By seeing if the heterodyne frequency increases or decreases, we can determine if the signal frequency is larger or smaller than the reference frequency.
As the measurements presented in Fig. \ref{fig3}d that $T_\phi=50 \unit{\mu s}$ is offset by a difference of 4 ns and $T_{seq}=10T_\phi$ changes accordingly, the detected heterodyne frequency of the signal shift in two different directions.
We further investigated the dependency of the heterodyne frequency on the parameters by performing measurements that vary (i) $T_{seq}$, (ii) $T_{seq}$ and $\omega_{ref}$.
When $T_\phi$ keeps unchanged, the heterodyne frequency shifts by
\begin{equation}
	\Delta\omega_H = \omega_{ref}\Delta T_{seq}/T_{seq}.
\end{equation}
Using the equation, we can estimate the frequency fidelity of the given sequence.
For example, with a timing error $<3$ ps, the frequency fidelity of a signal around 10 kHz could be only 0.06 mHz.
The frequency resolution can be arbitrarily high with a long $T_{seq}$ at the cost of bandwidth.

\subsection{Sensing of Arbitrary Audio Signals}

\begin{figure}[b]
	\includegraphics[scale=0.95]{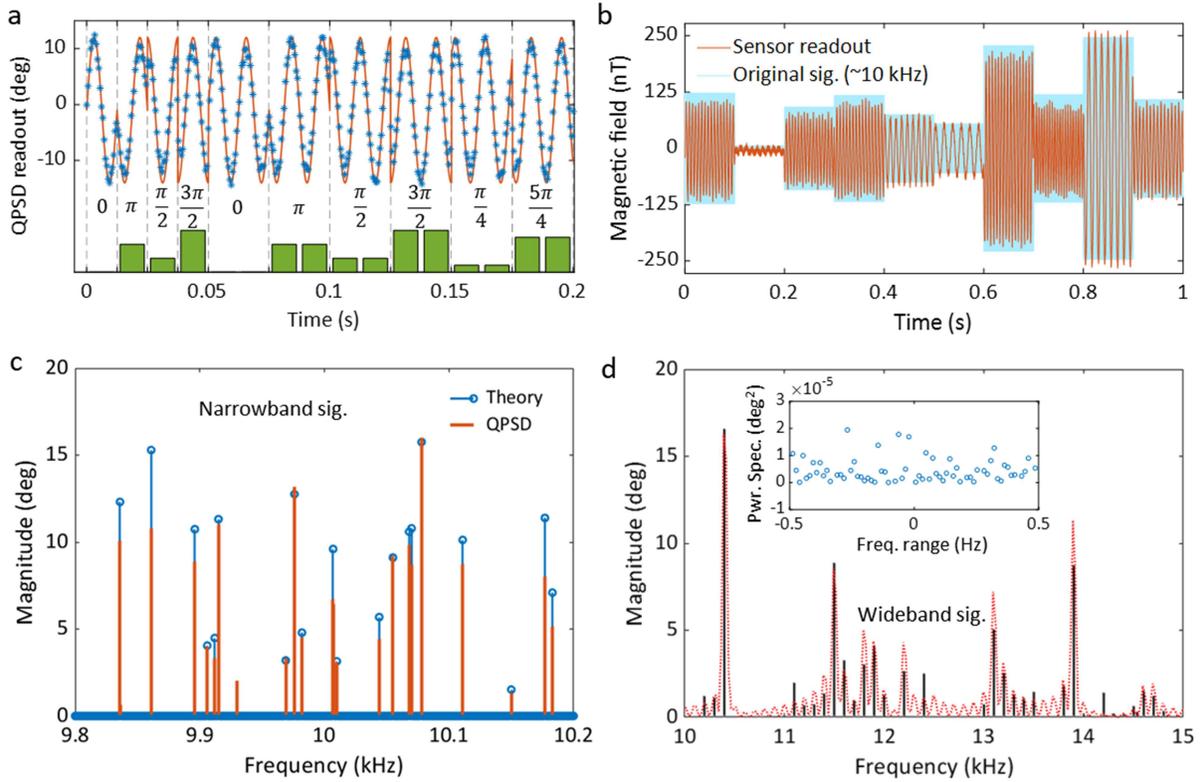}
	\caption{\label{fig4} Detection of arbitrary audio signals.
	(A) Phase response of the QPSD measurement.
	A 20.08 kHz signal with sequential phase changing is applied to the sensor.
	The bars show the phases at different time.
	Stars mark the readout of the sensor, and the curve is the simulated readout.
	(B) An ac field is applied with the frequency, amplitude and phase switched every 100 ms. The light blue areas corresponding to the right $y$-axis shows the applied field of around 10 kHz, and the red curve shows the QPSD readout.
	(C)	Spectral comparison of the applied signal and the detected magnetic field in a narrow bandwidth.
	The applied signal is a sum of 20 different sine signals within 400 Hz bandwidth.
	(D) A signal with wide bandwidth between 10 to 15 kHz is applied and detected by varying the sequence.
	We set an 800 Hz bandwidth for the measurement of each sequence and use 6 measurements to cover the entire bandwidth.
	The red dash line shows the spectrum of the output of the AWG, and the solid black line is the spectrum of the detected magnetic field signal.
	The inset figure is the phase-noise power spectrum density plotted within 1 Hz bandwidth cut.
	}
\end{figure}

We demonstrate measurements of arbitrary audio signals by combining the QPSD readout and the frequency offset heterodyne detection.
We first generate a signal at 20.08 kHz with its phase varying with time (see Fig. \ref{fig4}a). The MW filter is set by the Hahn-echo sequence with $T_\phi=50 \unit{\mu s}$.
With the reference frequency is at 20 kHz, the heterodyne readout is at 80 Hz, as seen from the simulated curve.
The phase of the external field is switched with a cycle of 80 Hz and 40 Hz so that the experimental readout displays the phase change, as shown in Fig. \ref{fig4}a.

Next, we apply a field with its frequency, amplitude and phase all arbitrarily changing.
The signal frequency is around 10 kHz and the signal bandwidth is within 400 Hz.
Using $1/T_{seq}=10 \unit{kHz}$, we can measure the signals close to 10 kHz with the same sensitivity as the 20 kHz signal.
The signal length is one second and consists of ten 100 ms parts.
In Fig. \ref{fig4}b, both the applied field waveform and the QPSD readout are plotted.
The heterodyne frequencies well resolve the frequency differences in the original waveform.
The amplitudes of the readout also correspond to the applied field strength.

As discussed previously, the measurement bandwidth used in the experiment is 400 Hz.
For this, we perform a spectrum analysis as shown in Fig. \ref{fig4}c.
The signal to be measured is a sum of 20 tones with random frequencies, amplitudes and phases. In order to distinguish the sign of frequency offsets for each component, we measure the signal using an alternative sequence with $T_\phi=50 \unit{\mu s}+2 \unit{ns}$.
The sharp peaks observed in the spectrum should shift according to the changes of the measurement sequence, else we exclude them as noise signals generated from our electronics.
As shown in Fig. \ref{fig4}c, the applied frequencies are properly resolved.
Additionally, we find a 9.93 kHz noise spike from the environment.
The errors in magnitude of the measured signal could be induced by the LIA filter, as shown earlier in Fig. \ref{fig2}d.
The errors could also be caused by an insufficient sampling number for demodulating the rotating frame modulation.
In the measurements, we apply sequences with their lengths corresponding to a sampling frequency $f_s=5 \unit{kHz}$.
The frequency difference of the two MWs is $\delta f=500$ Hz and $N=10$ for reading out a phase sample.
To increase the measurement precision, if we use a smaller $\delta f$, it will decrease the bandwidth.

Though the bandwidth of each measurement sequence is limited, we can still measure a signal with a wider bandwidth by merging several measurements.
The condition is that the signal bandwidth should not be larger than the sampling frequency to avoid frequency aliasing.
In Fig. \ref{fig4}d, we perform a spectrum analysis for a signal within a bandwidth from 10 kHz to 15 kHz.
The signal to be detected is a sum of 10 components with their frequencies randomly distributed in the bandwidth.
The signal is generated by an arbitrary signal generator (AWG) and sent to the test field coil. The dotted curve in the figure displays the spectrum of the electrical signal from the AWG. There are some harmonics near each main component due to the limited AWG internal clock and signal length.
The components at different frequencies are measured by varying $T_\phi$ to get different referencing frequencies for heterodyne detection.
The inset figure shows the power spectrum of the QPSD readout noise within 1 Hz bandwidth, from which we calculate the square root of the standard deviation $\sigma_{phase}=0.0022^\circ$.
The magnetic field sensitivity depends on the applied sequence and the corresponding frequency response.
In the case of the Hahn-echo sequence with $T_\phi=50 \unit{\mu s}$, we have a calibrated scalar factor of $k_{sf}=0.071^\circ/\unit{nT}$, and the magnetic field sensitivity is $\eta=31 \unit{pT/\sqrt{Hz}}$.
Taking the scalar factor into the calculation of LDR $[-\pi, \pi]$, we can get the dynamic range in decibels as $20\log[\pi/(k_{sf}\eta)]=98 \unit{dB}$.
The sensitivity can be further optimized by, e.g., using higher laser power, applying high-order dynamical decoupling sequences, and implementing flux concentration.

Finally, we demonstrate the detection and demodulation of audio signals.
Although 10 kHz is within the audio frequency band, most of the daily audio sounds have frequencies ranging from hundreds of Hz to a few kHz.
Therefore, signals need to be modulated to a detectable frequency range.
For this, we have used (i) a melody piece composed of 3 tones and (ii) a small part from Dr. Martin Luther King Jr.'s famous speech "I have a dream", to test the waveform reconstruction by the diamond quantum sensor.
The tones of the melody have frequencies distributed between 500 Hz and 700 Hz.
Therefore, we mix it with a 9.5 kHz reference to get the signal modulated around 10 kHz and broadcast the mixed signal to the diamond.
For case (ii), we have to compress the signal bandwidth into 200 Hz and then modulate it with a 10 kHz reference.
The audio reconstructed from the diamond sensor can be heard and compared with the original audio.
We have provided additional audio files in the Supplementary Materials confirming the same.

\section{Discussions}
In this work, we overcome the LDR limitation of the conventional interferometric readout through a new technique that includes the QPSD scheme and the frequency offset heterodyne readout.
The technique allows one to measure unknown signals with maximal sensitivity independent of their dynamic range.
It improves the feasibility for quantum sensors to perform high-sensitive measurements of different physical quantities using interferometric methods, beyond magnetometry.

Theoretically, the extended LDR comes from the multiple measurements that have the quantum phase evolving through the entire phase range $[-\pi, \pi]$ so that the initial phase factor that contains the external field information can be resolved.
Such an extended phase range affects the measurement bandwidth as well as the sensitivity.
In theory, the sensitivity does not deteriorate a lot from the conventional fluorescence readout except for a factor of $\sqrt{2}$.
While in the experiment, we suffer from a low contrast $C=0.19\%$ due to the low excitation laser power (80 mW).
The contrast and the fluorescence photon count can significantly increase when the laser reaches saturation power \cite{RN37}.
Different dynamical decoupling sequences can also improve the magnetic field sensitivity through the filter function $G(\omega)$.
Flux concentration could further improve the signal-noise ratio \cite{RN29,RN38}.
The flux concentrator can be very small compared to conventional dipole antennas because the gain no longer depends on the signal wavelength.
With the millimeter size diamond dimension, the flux concentration factor can easily reach a factor of hundreds when using a concentrator in centimeters.

The QPSD readout can also enhance the capability of vector magnetometry.
Conventionally, fluorescence emitted from NV centers in multiple orientations is measured sequentially to acquire the vector components.
Similar to the methods developed here, one could also modulate the signal on each orientation with different modulation frequencies \cite{RN39}.
Performing measurements on different NV orientations with appropriate synchronization can suppress the phase errors in vector reconstruction.

In conclusion, we demonstrated high sensitive detections of arbitrary magnetic field signals, including audio-range frequency signals like melody and speech, using the QPSD scheme in combination with the heterodyne readout.
A further improvement in sensitivity can be achieved by using flux concentrators.
One could also generalize the current methods to achieve vector magnetometry with extended LDR. We envisage that the techniques developed here will have the potential to develop low-distortion, small-volume quantum sensors for various applications in science and technology.

\section{Methods}
\subsection{Experimental setup}
The diamond used in the experiment is a (111)-oriented (0.5 mm)$^3$ cube obtained from a single crystal grown by the temperature gradient method at high-pressure high-temperature (HPHT) conditions.
The diamond is $99.97\%$ $^{12}$C enriched, and has an initial nitrogen concentration of 1.4 ppm.
The final NV concentration is 0.4 ppm after electron irradiation and annealing.
Dephasing time of the NV ensemble is obtained as $T_2^*=8.5 \unit{\mu s}$ by Ramsey sequence, and a decoherence time $T_2 = 200 \unit{\mu s}$ is measured by Hahn-echo sequence.
The diamond is positioned at center of a home-built three dimension coils system, and is illuminated by a 532 nm laser (Lighthous Sprout-G) at around 80 mW.
Microwave signals are generated from two sources (Rohde\&Schwarz, SMIQ03B) and are individually cut by two switches before the combination.
Measurement sequences are generated by a data timing generator (Tekreonix, DTG5274).
After the combination and amplification of the MW signals, MW pulses are feed to the diamond through a dielectric resonator antenna \cite{RN40}.
The detected fluorescence signal is demodulated by a LIA (Zurich Instruments, HF2LI) which has two independent differential input channels and demodulators.
To generate arbitrary magnetic fields, we write signals to an AWG (Tektronix, AWG520) with $10^5$ samples per second output sampling rate. The test signals are continuously repeated and sent to a copper loop near the diamond. 

\subsection{Spectrum analyzing}
The spectrum to be analyzed is divided into several sections with the bandwidth set by the LIA for data acquisition.
In each section, the center frequency determines $T_\phi$ of the measurement sequence.
Usually, the center frequency satisfies $f_c = 1/T_\phi+\varepsilon/T_{seq}$, where $\varepsilon = 0, \pm1$.
A time trace is recorded after running the sequence, and a spectrum is acquired from the Fourier transform of the time trace.
However, the spectrum is a fold of the two sidebands with repect to the center frequency.
The sequence with $T_\phi'=T_\phi+t_{clk}$ and $T_{seq}'=mT_\phi'$ is applied to get an alternate spectrum with analyzed frequencies shift by $\Delta f=\pm\left|1/T_\phi-1/T_\phi'\right|$.
The direction of the frequency shift shows which side the signal component belongs to.
In the algorithm, we set a threshold to separate signal spikes from noise, and use the known sequences induced spectrum frequency shift to distinguish the signs of the signal offset frequency to the center frequency.
The signal spikes that do not shift according $\Delta f$ are recognized as systematic noise spikes.
Then, the spectrum of the selected section can be replotted as the example shown in Fig. \ref{fig4}c.
After measuring the spectra of all the sections, we can get the final spectrum by merging them together.
	
\section{Acknowledgments}
We acknowledge financial support by European Union’s Horizon 2020 research and innovation program ASTERIQS under grant No. 820394, European Research Council advanced grant No. 742610, SMel, Federal Ministry of Education and Research (BMBF) project MiLiQuant and Quamapolis,
German Research Foundation grant GRK 2198 and 2642, and Japan Society for the Promotion of Science (JSPS) KAKENHI No. 17H02751.

\nocite{*}
	

%

\clearpage
\section*{Supplemental information: Low Distortion Radio Signal Sensing with a Quantum Sensor}
\renewcommand{\theequation}{S\arabic{equation}} 
\renewcommand{\figurename}{S-FIG.} 
\setcounter{figure}{0}
\setcounter{equation}{0}

\subsection{Derivation of the rotating frame modulation}
Based on Eq. (2) in the main text, the evolution operator with $\pi/2$ pulse duration $\tau_1=\pi/(2\Omega_1)$ is
\begin{equation}
	e^{-i\frac{\Ham_1'}{\hbar}t}=\frac{1}{\sqrt{2}}
	\begin{pmatrix}
		1 & -ie^{-i\left(\delta\omega_1 \tau_1 + \alpha\right)}\\
		-i e^{i\left(\delta\omega_1 \tau_1 + \alpha\right)} & 1
	\end{pmatrix}.
\end{equation}
In the sensing duration, the spin state acquires an extra phase factor due to the external field and can be expressed as
\begin{equation}
	\ket{\psi(T_\phi)} = \frac{1}{\sqrt{2}}
	\begin{pmatrix}
		-ie^{-i(\delta\omega_1 \tau_1+\alpha +\phi)}\\
		1
	\end{pmatrix}.
\end{equation}
The evolution operator during the MW2 pulse is similar to (S1), and the spin state after the second $\pi/2$ pulse is
\begin{equation}
	\ket{\psi_{meas}} = \frac{1}{\sqrt{2}}
	\begin{pmatrix}
		e^{-i\Phi}-ie^{-i\delta\omega_2\tau_2 +\beta)}\\
		-i e^{-i\Phi} e^{i(\delta\omega_2\tau_2 +\beta)}+1
	\end{pmatrix}.
\end{equation}
where $\Phi=\delta\omega_1\tau_1+\alpha+\phi+\pi/2$, $\tau_2=\pi/(2\Omega_2)$ is the $\pi/2$ pulse duration of MW2. 
Then, we can calculate the measured expectation value as
\begin{equation}
	\braket{S_z} = \frac{\hbar}{2}\bra{\Psi_{meas}}\sigma_z\ket{\Psi_{meas}}= \sin\left[\phi+\frac{\pi}{2}\left(\frac{\delta\omega_1}{\Omega_1}-\frac{\delta\omega_2}{\Omega_2}\right)+\alpha-\beta\right],
\end{equation}
where $\alpha-\beta=\delta\omega\cdot t$, $\delta\omega=\delta\omega_1-\delta\omega_2=2\pi\delta f$

\subsection{Microwave filter function}
Here we give a general switching function based on CPMG-n sequence.
Typically, a microwave sequence consisting of $\pi$-pulses can be described by a switching function
\begin{equation}
	g(t)= \left\{
	\begin{matrix}
		1,	&	t\in\left[0, \displaystyle{\frac{T_\phi}{2n}}\right)\\
		(-1)^k,	&	t\in\left[\displaystyle{\frac{2k-1}{2n}}T_\phi, \displaystyle{\frac{2k+1}{2n}}T_\phi\right)\\
		(-1)^n,	&	t\in\left[\displaystyle{\frac{2n-1}{2n}}T_\phi, T_\phi\right)\\
		0,	&	t\in\left[T_\phi, mT_\phi\right]\\
	\end{matrix}\right.,
\end{equation}
where $k\le n-1$ and $k, n, m \in \mathbb{N}$.The switching function is repeated in each time interval of $\left[NmT_\phi,(N+1)mT_\phi\right]$, where $N\in\mathbb{Z}$. Given a random sinusoid signal $B_{ac}(t)=Be^{-i(\omega t+\varphi)}$, the readout of each cycle is given by $\phi_r(t)=\int_{-\infty}^{\infty}s(t)g(t)dt$. For each sampling point, the accumulated phase factor is
\begin{equation}
	\phi_r(N)=\left\{
	\begin{matrix}
		\displaystyle{\frac{4\left(\sin \frac{\omega T_\phi}{4n}\right)^2 \left(\cos \frac{\omega T_\phi}{2}\right)}{\omega \cos \frac{\omega T_\phi}{2n}}} e^{i(-\frac{\omega T_\phi}{2}-\frac{\pi}{2})} Be^{-i\varphi}e^{-i \omega NmT_\phi},	&n\ is\ odd;\\
		\\
		\displaystyle{\frac{4\left(\sin \frac{\omega T_\phi}{4n}\right)^2 \left(\sin \frac{\omega T_\phi}{2}\right)}{\omega \cos \frac{\omega T_\phi}{2n}}} e^{i(-\frac{\omega T_\phi}{2}-\pi)} Be^{-i\varphi}e^{-i \omega NmT_\phi},	&n\ is\ even.\\
	\end{matrix}\right.
	\label{eq_acResp}
\end{equation}
The equation is simplified as
\begin{equation}
	\phi_r(N)=|G_n(\omega)|e^{i\left(-\frac{\omega T_\phi}{2}-\frac{P}{2}\pi\right)}\gamma_eB(\omega)e^{-i\varphi(\omega)}e^{-i\omega NmT_\phi},
\end{equation}
where $G_n(\omega)=|G_n(\omega)|e^{i\left(-\frac{\omega T_\phi}{2}-\frac{P}{2}\pi\right)}$ is the MW filter function of the sequence. When $n=1$ and 2, i.e., Hahn-echo and CPMG-2 sequence, the sensor responses are
\begin{equation}
	G_1(\omega)=\displaystyle{\frac{4}{\omega}\left(\sin{\frac{\omega T_\phi}{4}}\right)^2} e^{-i\left(\frac{\omega T_\phi}{2}-\frac{\pi}{2}\right)}
	\label{eq_Hahn}
\end{equation}
\begin{equation}
	G_2(\omega)=\displaystyle{\frac{8}{\omega} \left(\sin{\frac{\omega T_\phi}{8}}\right)^2\left(\sin{\frac{\omega T_\phi}{4}}\right)e^{-i\left(\frac{\omega T_\phi}{2}-\pi\right)}}
	\label{eq_CPMG2}.
\end{equation}
When $\omega=2\pi/T_\phi$, the two sequences have the equal response as $|G_1|=|G_2|=2T_\phi/\pi$.
The phase response difference is $\pi/2$. Therefore, the Hahn-echo sequence can be used to measure the $X(\omega, t)$ component for the quantum-LIA, in correspondence to a classical LIA. The CPMG-2 sequence can be used to measure the $Y(\omega, t)$ component. Furthermore, we can get $|G_1(2\pi/T_\phi)| = |G1(\pi/T_\phi)|$ from (S8).

\subsection{Derivation of shot-noise limited sensitivity}
According to the schematic described in the main text, we try to extract the phase factor from the N samples in a modulation cycle.
The shot-noise contribution to each sample is $\delta\mathcal{F}=\sqrt{2\mathcal{N}}$, where $\mathcal{N}$ is the detected photon number in each measurement, and there is a factor of $\sqrt{2}$ because each sample include two measurements.
The output from the photodetector can be described as
\begin{equation}
	s_k=\mathcal{N}C\sin\left(\frac{2k\pi}{N}+\phi\right)
	\label{sk},
\end{equation}
where $C$ is the signal contrast and $k$ denotes the number in the samples.
Since the photodetector output include the shot-noise, the measured $\overline{s}_k$ include the noise $\delta\overline{s}_k\approx\delta\mathcal{F}$.
We can define a minimum error square as
\begin{equation}
	\chi^2 = \sum_{k=1}^{N}(\overline{s}_k-s_k)^2.
\end{equation}
Additionally, we have $\sum_{k=1}^{N}\sin^2(2k\pi/N+\phi)=N/2$, and $\sum_{k=1}^{N}\sin(2k\pi/N+\phi)\cos(2k\pi/N+\phi)=0$ when the each modulation cycle is equally sampled with an even number.
Let $\partial\chi^2/\partial\phi=0$ to minimize $\chi^2$,
\begin{equation}
	\sum_{k=1}^{N}\overline{s}_k\cos\left(\frac{2k\pi}{N}+\phi\right)=0
\end{equation}
Take a differential,
\begin{equation}
	\sum_{k=1}^{N}\left[\delta\overline{s}_k\cos\left(\frac{2k\pi}{N}+\phi\right)-\overline{s}_k\sin\left(\frac{2k\pi}{N}+\phi\right)\delta\phi\right]=0
\end{equation}
Sample $\overline{s}_k$ can be replaced by $s_k+\delta\overline{s}_k$. We can neglect the high order error $\delta\overline{s}_k\delta\phi$ and get
\begin{equation}
	\sum_{k=1}^{N}\delta\overline{s}_k\cos\left(\frac{2k\pi}{N}+\phi\right)=\delta\phi\sum_{k=1}^{N}\mathcal{N}C\sin^2\left(\frac{2k\pi}{N}+\phi\right)=\mathcal{N}C\frac{N}{2}\delta\phi
\end{equation}
The statistical average of the phase variances square is
\begin{eqnarray}
	\braket{(\delta\phi)^2}=\frac{4}{N^2}\frac{1}{(\mathcal{N}C)^2}\sum_{k=1}^{N}\sum_{l=1}^{N}\braket{\delta\overline{s}_k\delta\overline{s}_l}\cos\left(\frac{2k\pi}{N}+\phi\right)\cos\left(\frac{2l\pi}{N}+\phi\right)
\end{eqnarray}
The measurement errors should be uncorrelated, satisfying $\braket{\delta\overline{s}_k\delta\overline{s}_l}=0$.
Therefore,
\begin{equation}
	\braket{(\delta\phi)^2}=\frac{4}{N^2}\frac{1}{(\mathcal{N}C)^2}\sum_{k=1}^{N}\braket{(\delta\overline{s}_k)^2}\cos^2\left(\frac{2k\pi}{N}+\phi\right)=\frac{2}{N}\frac{1}{(\mathcal{N}C)^2}\sigma_F^2.
\end{equation}
The short-noise determined phase noise in the modulation cycle $2NT_{seq}$ is
\begin{equation}
	\delta\phi=\sqrt{\frac{2}{N}}\frac{\delta\mathcal{F}}{\mathcal{N}C}=\frac{2}{\sqrt{N}}\frac{1}{C\sqrt{\mathcal{N}}}.
\end{equation}
Taking $\delta\phi=\gamma_e G(\omega)\delta B$ into the equation, we can get a sensitivity
\begin{equation}
	\eta_{Hahn}=\frac{\sqrt{2}\pi}{\gamma_e C\sqrt{\mathcal{N}}}\frac{\sqrt{T_{seq}}}{T_\phi}.
\end{equation}
In order to compare with the sensitivity of the direct fluorescence readout, we can get the minimum detectable phase when the signal satisfy the small-angle approximation as
\begin{equation}
	\delta\phi'=\frac{\delta\mathcal{F}}{C\mathcal{N}}=\frac{\sqrt{2}}{C\sqrt{\mathcal{N}}}
\end{equation}
Therefore, the sensitivity of Hahn-echo measurement with fluorescence readout will be
\begin{equation}
	\eta_{Hahn}' = \frac{\pi}{\gamma_e C \sqrt{\mathcal{N}}}\frac{\sqrt{T_{seq}}}{T_\phi}.
\end{equation}
The shot-noise limited sensitivity of the QPSD readout deteriorates from the value of the fluorescence readout by a factor of $\sqrt{2}$.

Taking the values in experiments, that $\mathcal{N}=4.6\times10^{11}$, $C = 0.19\%$, $T_\phi=50 \unit{\mu s}$, $T_{seq}=100 \unit{\mu s}$, and with $\gamma_e=2\pi\times(28 \unit{Hz/nT})$, we calculate that the shot-noise limited sensitivity using QPSD readout is $\eta\approx4 \unit{pT/Hz^{1/2}}$.

\subsection{Frequency resolution}
The following is the detailed description of the measurements of Fig. 3d that discusses the frequency resolution of the experiment.
Firstly, we apply sequences with $T_\phi = 50 \unit{\mu s}$, $50 \unit{\mu s}\pm4\unit{ns}$ and $T_{seq} = 10T_{\phi}$ to measure a signal at 20.005 kHz, denoted as Meas. 1 in the figure.
The reference frequency shifts 1.6 Hz due to the changing of $1/T_\phi$.
As a result, a heterodyne frequency of 6.6 Hz can be detected by measuring a 20.005 kHz field with $T_\phi=50 \unit{\mu s}-4 \unit{ns}$.
Correspondingly, a 3.4 Hz signal will be detected when measuring a 19.995 kHz field so that it becomes distinguishable to the 20.005 kHz field.
For fields at frequencies near the other references $\omega_{ref}$, the shifts can also be calculated accordingly by $k/(mT_\phi)$.
To further understand how sensitive is $\omega_{ref}$ to the sequence parameters, we only differ $\Delta T_{seq}=\pm4$ns to measure external fields at 20.005 kHz and 16.005 kHz while keeping $T_\phi =50 \unit{\mu s}$.
In this case, the heterodyne signal is introduced only by the accumulated phase error in each cycle.
The reference frequency shift is calculated by Eq. (7) in the main text.
For example, by measuring a signal around 16 kHz with $\Delta T_{seq}=4$ ns, we detect a heterodyne frequency shift of 128 mHz as shown in Fig. 3d.
Therefore, the frequency fidelity is majorly limited by the jitter performance of the pulse generator.
We calculate an example that has a timing error $<3$ ps, according to the datasheet of the pulse generator we use (Tektronics, DTG5274).
The frequency fidelity is 0.06 mHz when the measured signal is around 10 kHz.

\subsection{Audio signal measurement}
The first file is a piece of melody that we composed with tones C5, D5, and E5.
The corresponding frequencies are $f_1=523$Hz, $f_2=587$Hz, $f_3=659$Hz. The signal waveform plotted in the time domain can be seen in Fig. S1.
We choose the three tones with their frequency differences within 200 Hz because the used quantum phase modulation frequency is 500 Hz and the LIA limited bandwidth is 200 Hz.
The signal is modulated with a carrier of 9.5 kHz so that the transmitted magnetic field is at frequencies around 10 kHz, which can be detected by the sensor using the Hahn-echo sequence with $T_\phi=50 \unit{\mu s}$.
Then, the QPSD readouts of the three tones will be at frequencies $f_{r1}=23$Hz, $f_{r2}=87$Hz, $f_{r3}=159$Hz.
In order to reconstruct the original signal, the readout is mixed with a 500 Hz carrier.
Three notch filters are applied to the mixed signal to remove the sideband as the final step.

As for the speech signal, the signal is distributed over a frequency range from DC to a few kHz. We cut off the spectrum from 4 kHz, shown in Fig. S2(a), because signal amplitude after 4 kHz is much smaller than the components at lower frequencies.
In order to make the signal detectable by the sensor, the signal is firstly compressed in the frequency domain by using interpolation while keeping the same sampling rate for broadcasting. The signal size after interpolation is 20 times the original signal size so that in the frequency domain, the bandwidth is compressed from 4 kHz to 200 Hz, as shown in Fig. S2(b). Then, the signal is mixed with a 10 kHz reference so that we can use the same Hahn-echo sequence to measure.
Figure S2(C) shows the spectrum in which the audio signal spectrum distributes in symmetric to 10 kHz.
Finally, the QPSD readout detects the heterodyne signal, of which the frequency is referenced to 10 kHz.
To reconstruct the original sound, we compress the readout in time domain, i.e., averaging the readout by every 20 data points.
In the frequency domain, the spectrum bandwidth is expanded from 200 Hz back to 4 kHz.
\pagebreak
\begin{figure}[h]
	\includegraphics[scale=1.25]{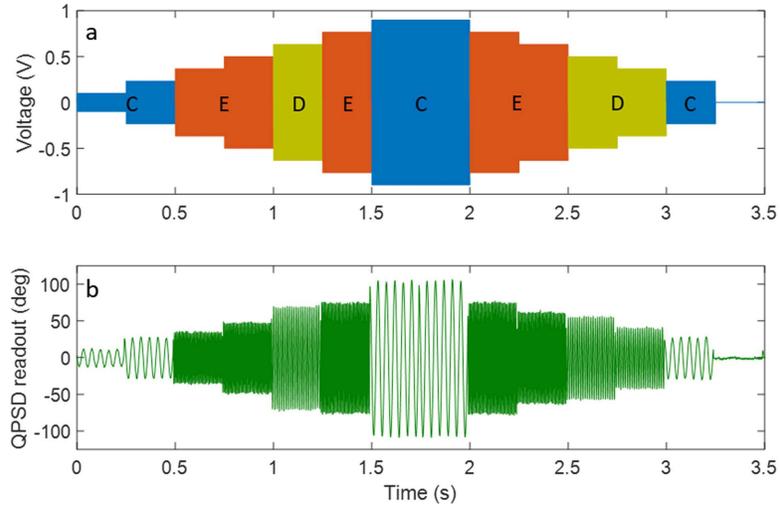}
	\caption{\label{FigS1} (A) Signal waveform of the melody played for the detection. The voltage signal is sent to an AWG for generating the magnetic field through a loop antenna. The different colors represent different scales of the tones that are labeled in the figure. Changing the amplitude indicates that the loudness of the sound rises at the beginning and then falls in the end. (B) The QPSD readout of the sensor. Different tones can be recognized from the different heterodyne frequencies in the readout.}
\end{figure}
	
\begin{figure}[t]
	\includegraphics[scale=0.95]{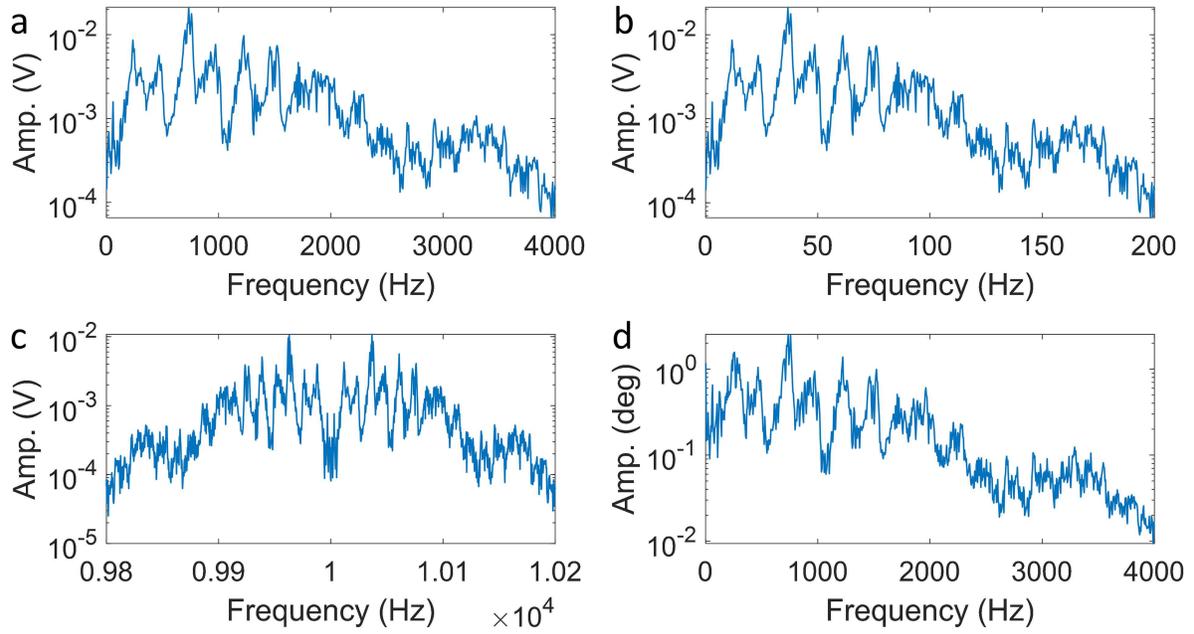}
	\caption{\label{FigS2} (A) Spectrum of the original speech signal. The signal is written into an AWG in the unit of Volt. Therefore, the amplitudes of the spectra in sub-figure A to C are all in Volt. (B) Frequency compression. The signal is interpolated to a data size of 20 times larger than the original one. Therefore, the signal is compressed in the frequency domain. (C) The signal spectrum after mixing with a 10 kHz reference signal. (D) The QPSD readout spectrum. The heterodyne signal is sampled every 20 data from the raw readout to reconstruct the audio signal spectrum.}
\end{figure}

\end{document}